\documentclass[reprint]{revtex4-1}
\usepackage{pifont}
\usepackage{graphicx,colordvi}
\usepackage{amsmath,amssymb,amsthm}
\usepackage[colorlinks=true,urlcolor=blue,linkcolor=blue,citecolor=red]{hyperref}

\newcommand{\ini}{\begin{equation}}
\newcommand{\fin}{\end{equation}}
\newcommand{\inia}{\begin{eqnarray}}
\newcommand{\fina}{\end{eqnarray}}

\begin{document}

\title{\bf Newtonian gravity and resonance on de-Sitter branes}
\author{Adriana Araujo $^1$, Rommel Guerrero $^2$ and R. Omar Rodriguez $^2$}
\address{ $^1$Departamento de Matem\'atica and   $^2$Departamento de F\'isica, Universidad Centroccidental Lisandro Alvarado, Barquisimeto,Venezuela}

\begin{abstract}

A dS brane on the boundary  between two five-dimensional spacetimes is determined. We consider asymmetric scenarios with AdS${}_5$ vacua at each side of the dS brane; and as a result, a resonant mode inside of the spectrum of the gravitational fluctuations is found.  We analyze the deviations to the Newton potential generated by the gravitational excitations, finding that, for scenarios with large values of the cosmological constants,  the contribution of the resonant mode is exponentially suppressed. However, when  one of the vacua is null, the resonant mode belongs to the light states set of the gravitational fluctuations and  five-dimensional gravity is recovered on the dS brane.

\vspace{0.5 cm} 
\end{abstract}

\maketitle

\section{Introduction}

In  theories with extra dimensions  our Universe is conceived inside of a geometrical structure with more than four dimensions. In this context, we find the brane world models, where the fundamental interaction fields  are confined on a brane in the four-dimensional  transition region between two spacetimes with different curvatures.  In particular, the gravitational fluctuations are characterized by a ground state localized around the brane and by a massive tower of unbounded states which generate corrections to the four-dimensional term in the gravitational  potential \cite{Randall:1999vf}.

These configurations can be determined either as  vacuum solutions to the Einstein equations rigidly connected on the four-dimensional  transition region \cite{Cvetic:1993xe}; or as self-gravitating solutions to the coupled Einstein-scalar field system,  where the scalar field is a  topological kink interpolating between the minima of a
potential with spontaneously broken symmetry.  The energy density of these solutions is associated to a wall which divides the spacetime into two portions whose curvatures are identified with the extrema of the scalar potential. Then,  in the thin wall limit of the self-gravitating wall the brane is obtained  \cite{Guerrero:2002ki}.

In both cases, it is possible to consider de-Sitter (dS) scenarios with metrics similar to the Friedmann-Robertson-Walker solution \cite{Goetz:1990,Wang:2002pka,Kehagias:2002qk,Ghoroku:2003bs}, where the dS brane, unlike the static cases, could be the boundary  between two flat spaces or, in more general cases,  with positive curvatures. However, only when the vacua are negatives or Anti de-Sitter (AdS) the standard four-dimensional potential is recovered on the brane \cite{Kehagias:2002qk,Ghoroku:2003bs}. On the other hand, asymmetric dS scenarios where the brane interpolates between different vacua; e.g., dS-AdS, such that the gravity is confined in an unusual way, have been studied in \cite{Guerrero:2005xx,Guerrero:2005aw}. Nevertheless, the corrections to the four-dimensional term in the gravitational potential have not yet been analyzed; and it is interesting to determine if only a negative vacuum is sufficient to recover Newtonian gravity on any four-dimensional sector of the asymmetric scenario considered there.

In order to determinate the deviations to the Newtonian potential in an asymmetric scenario, the existence of a resonant mode  inside of the spectrum of gravitational fluctuations must be considered. In the static case,  its contribution to the gravitational  potential on the brane was  first analyzed in \cite{Gabadadze:2006jm}, and more recently in \cite{Melfo:2010xu}. On the other hand, in the dynamics case the analysis has not yet been done; and it is our intention to address this issue throughout this paper. For this, we will  identify  a set of integrability conditions for the gravitational fluctuations similar to those found in \cite{Melfo:2010xu}.

The paper is organized as follows. In next Section, we shall find a  five-dimensional family of dS vacuum solutions to the Einstein equations with arbitrary cosmological constants at each side of the brane. In Section \ref{GraRes}, the corrections to the Newtonian potential  generated   by the massive tower of the gravitational excitations will be determined, considering several scenarios with AdS${}_5$  vacua and taking into account  the possible effects produced by the resonant modes. Finally, we will present our conclusions in Section \ref{diss}.

\section{The scenario}

We consider a five-dimensional structure, where $z$ is the extra coordinate, generated by two spacetimes with arbitrary curvatures  connected rigidly  on  the hypersurface $z=0$. The metric tensor of the scenario is given by
\begin{equation}
 g_{ab}=\Theta(-z)g_{-ab}+\Theta(z)g_{+ab},\quad a, b=0,\dots ,4,  \label{Metric}
\end{equation}
such that 
\begin{equation}
g_{-ab}(0)=g_{+ab}(0)
\end{equation}
which can be determined as  a three-brane solution to the Einstein field equations 
\begin{equation}
 G_{ab}-\Lambda g_{ab}=T_{ab},
 \end{equation}
 with
  \begin{equation}
\Lambda=\Theta(-z)\Lambda_- +\Theta(z)\Lambda_+ \label{CtteCosml}
\end{equation}
and
 \begin{equation}
 T_{ab}=-\tau\delta(z) g_{\alpha\beta}\delta_{a}^{\alpha}\delta_{b}^{\beta},\quad \tau>0,\quad \alpha, \beta=0,\dots ,3
 \end{equation}
where $\tau$ is the brane tension. 

In particular, consider the embedding of a thin dS brane into a five-dimensional bulk described by a metric with planar-parallel symmetry, with associated line element 
\begin{equation}
ds^2=f^2(z)\left(-dt^2+dz^2+e^{2\beta t}\delta^{ij}dx_i dx_j\right),\; \beta>0,\label{g}
\end{equation}
where
 \begin{eqnarray}\label{f1}
 f^{-1}(z)&=&\nonumber\\
 &&\begin{cases}
      e^{\beta z}+ \left(1-\sqrt{1-\frac{\Lambda_-}{6\beta^2}} \right)\sinh\beta z,& z < 0 \cr\cr
      e^{-\beta z} -\left(1-\sqrt{1-\frac{\Lambda_+}{6\beta^2}} \right)\sinh\beta z,& z > 0, \cr
        \end{cases}
 \end{eqnarray}
and with tension given by
 \begin{equation}
 \tau=3\left(\sqrt{\beta^2-\Lambda_+/6} +\sqrt{\beta^2-\Lambda_-/6}\right),\; \Lambda_\pm\leq6\beta^2.  \label{TensionAsimt}
 \end{equation}

For the particular case
\begin{equation}
\Lambda_-=6\alpha (2\beta-\alpha),\;\;\Lambda_+=- 6\alpha (2\beta+\alpha),\;\; 0<\alpha<2\beta,\label{CtteCosml2}
\end{equation} 
where  the symmetry around the brane is defined by $\alpha$, the metric factor (\ref{f1}) is reduced to
\begin{equation}
f(z)=\left(e^{\beta |z|}+\frac{\alpha}{\beta}\sinh\beta z\right)^{-2}\ ,\quad 0<\alpha<2\beta\label{g0}\ ;
\end{equation}
and the metric tensor (\ref{g}, \ref{g0}) can be identified with the thin wall limit of the self-gravitating wall given by (\ref{g}) and
\begin{equation}
f(z)^{-1}=\cosh^\delta\frac{\beta z}{\delta}+i\text{sgn}(z)\frac{\alpha\delta F}{\beta(1-2\delta)}\cosh^{1-\delta}\frac{\beta z}{\delta},\label{f-1}
\end{equation}
where $F\equiv{}_2F_1\left[1/2-\delta,1/2,3/2-\delta,\cosh^2\beta z/\delta\right]$ is the hypergeometric function and $\delta$ is the thickness of the wall. The spacetime (\ref{g}, \ref{f-1}) was determined in \cite{Guerrero:2005aw}, it corresponds to a scenario where the zero mode of gravity fluctuations is confined on the four-dimensional transition region between the spaces with cosmological constants (\ref{CtteCosml2}).

Notice that in the null curvature limit, $\Lambda_\pm\rightarrow 0$, (\ref{f1}) is reduced to
 \begin{equation}
 f\rightarrow e^{-\beta|z|} ,
 \end{equation}  
which is the metric factor of the dS thin wall found in \cite{Guerrero:2002ki}, whose self-gravitanting version was reported initially in \cite{Goetz:1990}. On the other hand, in the static limit, $\beta\rightarrow 0$, when $\Lambda_\pm=\Lambda<0$, (\ref{g}, \ref{f1}) consistently converges to the RS-2 scenario \cite{Randall:1999vf} in conformal coordinates
\begin{equation}
f\rightarrow\left(1+\sqrt{|\Lambda|/6}\ z\right)^{-1}.
\end{equation} 

We will analyze the gravitational interaction on the four-dimensional sector of the spacetime (\ref{g}, \ref{f1}).

\subsection{Gravitational fluctuations}

In order to determinate the spectrum of the gravitational fluctuations of (\ref{g}, \ref{f1}), consider small perturbations, $h_{ab}$, in the transverse and traceless sector. Thus, in the axial gauge, $h_{az}=0$, and under the usual decomposition, $h_{\mu\nu}=e^{ip\cdot x}f^{1/2}(z)\psi_{\mu\nu}(z)$, with $m^2\equiv -p_\mu p^\mu-2\beta^2$, we have ($\mu\nu$ index omitted)
\begin{equation}
-\frac{d^2}{dz^2}\psi_m (z)+V_{QM}\psi_m(z)=m^2 \psi_m(z), \label{sch}
\end{equation}
where
\begin{eqnarray}
V_{QM}(z)&=&-\frac{\tau}{2} \delta(z)+\frac{9}{4} \beta^2\nonumber\\\nonumber\\
&&-15 \left(\frac{C_- \beta e^{\beta z}}{1+C_-^2 e^{2\beta z}}\right)^2 \Theta(-z)\nonumber\\\nonumber\\
&&-15\left(\frac{C_+ \beta e^{-\beta z}}{1+C_+^2e^{-2\beta z}}\right)^2 \Theta(z),\label{VQM}
\end{eqnarray}
and
\begin{equation}
C_{\pm}=\sqrt{\frac{\beta^2}{\Lambda_\pm/6}}-\sqrt{\frac{\beta^2}{\Lambda_\pm/6}-1}.
\end{equation}

The eigenfunctions of (\ref{sch}, \ref{VQM}) are determined by a zero mode localized around the brane,
$
\psi_0(z)=N_0 f^{3/2}(z)
$
 with
\begin{eqnarray}
N_0^{-2}&=&
-\frac{1}{2}\sqrt{\frac{6}{\Lambda_-}\left(\frac{6\beta^2}{\Lambda_-}-1\right)}+\beta^2\sqrt{\left(\frac{6}{\Lambda_-}\right)^3}\nonumber\\&&\times\tan^{-1}\left(\sqrt{\frac{6\beta^2}{\Lambda_-}}-\sqrt{\frac{6\beta^2}{\Lambda_-}-1}\right)\nonumber\\
&&-\frac{1}{2}\sqrt{\frac{6}{\Lambda_+}\left(\frac{6\beta^2}{\Lambda_+}-1\right)}+\beta^2\sqrt{\left(\frac{6}{\Lambda_+}\right)^3}\nonumber\\&&\times\tan^{-1}\left(\sqrt{\frac{6\beta^2}{\Lambda_+}}-\sqrt{\frac{6\beta^2}{\Lambda_+}-1}\right),\label{N0}
\end{eqnarray}
and a continuos tower of massive modes propagating freely for the five-dimensional bulk. By defining 
\begin{equation}
\mu\equiv\sqrt{m^2-9\beta^2/4}
\end{equation}
and
\begin{equation}
F_{\pm}\equiv {}_2F_1\left[\frac{5}{2}, -\frac{3}{2}; 1- i\frac{\mu}{\beta}; \left(1+ \frac{1}{C_\pm^2}e^{\pm 2\beta z}\right)^{-1}\right]
\end{equation}
we have, for $z<0$
\begin{eqnarray}
\psi_{m-}(z)&=&N_m\left[A_- \left(e^{-i \mu z}F_{-}+e^{i \mu z}F^*_{-}\right)\right.\nonumber\\&&\left. -i B_-\left(e^{-i \mu z}F_{-}-e^{ i \mu z}F^*_{-}\right)\right],\label{modosmasivos-}
\end{eqnarray}
and for $z>0$
\begin{eqnarray}
\psi_{m+}(z)&=&N_m\left[A_+ \left(e^{i \mu z}F_{+}+e^{- i \mu z}F^*_{+}\right)\right.\nonumber\\&&\left.- i\left(e^{i \mu z}F_{+}-e^{- i \mu z}F^*_{+}\right)\right],\label{modosmasivos+}
\end{eqnarray}
where $N_m$ and $A_\pm, B_-$ are the normalization and integration constants respectively, to be determined from the integrability conditions of the problem (\ref{sch}, \ref{VQM}).
 
Notice that, in correspondence with (\ref{VQM}), the massive modes are separated from the bound state by a mass gap of $9\beta^2/4$, which is a generic property of the dS scenarios. 

According with \cite{Melfo:2010xu}, any eigenvalue of  (\ref{sch}, \ref{VQM}) is associated to two wave functions identified as $\psi_m^c$ and $\psi_m^d$, which satisfy  
\begin{eqnarray}
&&\psi_{m-}^c(0)=\psi_{m+}^c(0)=0,\label{saltoc0}\\
&&\frac{d}{dz}\psi^{c}_{m-}(0)-\frac{d}{dz}\psi^{c}_{m+}(0)=0,\label{saltoc}
\end{eqnarray}
and
\begin{eqnarray}
&&\psi_{m-}^d(0)=\psi_{m+}^d(0),\label{saltod0}\\ 
&& \frac{d}{dz}\psi^{d}_{m-}(0) -\frac{d}{dz}\psi^{d}_{m+}(0)=\frac{\tau}{2}\psi_m^d(0),\label{saltod}
\end{eqnarray}
together with 
\begin{equation}
 \int_{-\infty}^{\infty} \psi^{*i}_{m^\prime}(z)\psi^j_m(z)dz=\delta^{i j}\delta(m-{m^\prime}),\quad i, j=c, d.\label{ortonormal}
\end{equation}

Thus, the four integration constants for $\psi_m^c$ and $\psi_m^d$, can be determinate by (\ref{saltoc0}, \ref{saltoc}) and (\ref{ortonormal}) in the first case; and by (\ref{saltod0}, \ref{saltod}) and (\ref{ortonormal}) in the second one.

Notice that the orthonormality condition (\ref{ortonormal}) is divergent for all $m=m^\prime$, which is a problem that can be solved by  introducing two regularity branes at $\pm z_r$ \cite{Callin:2004py}, such that  the initial scenario is recovered in the limit $z_r\rightarrow\infty$. 

As consequence of the regulatory branes, the potential (\ref{VQM}) is modified and now exhibits, besides  the well, two infinite height barriers, 
\begin{eqnarray}
V_{QM}(z)&=&-\frac{\tau}{2}\delta(z)+\frac{9}{4}\beta^2\nonumber\\\nonumber\\
&&+3\beta\left(\frac{1-C_-^2 e^{-2\beta z_r}}{1+C_-^2 e^{-2\beta z_r}}\right)\delta(z+z_r)\nonumber\\\nonumber\\
&&+3\beta\left(\frac{1-C_+^2 e^{-2\beta z_r}}{1+C_+^2 e^{-2\beta z_r}}\right)\delta(z-z_r)\nonumber\\ \nonumber\\
&&-15 \left(\frac{C_- \beta e^{\beta z}}{ 1+C_-^2 e^{2\beta z}}\right)^2\Theta(z+z_r) \Theta(-z)\nonumber\\ \nonumber\\
&&-15 \left(\frac{C_+ \beta e^{-\beta z}}{ 1+C_+^2 e^{-2\beta z}}\right)^2\Theta(z_r-z) \Theta(z), \label{VQMf2}
\end{eqnarray} 
and the massive modes  satisfy the following integration condition   
\begin{eqnarray}
\frac{d}{d z}\psi_\pm^d(\pm z_r)=\mp3\beta\left(\frac{1-C_\pm^2 e^{-2\beta z_r}}{1+C_\pm^2 e^{-2\beta z_r}}\right)\psi_\pm^d(\pm z_r),\label{cuantiz}
\end{eqnarray}
additional to (\ref{saltod}).  In the  limit $z_r\rightarrow\infty$, from (\ref{cuantiz}), is obtained that  the massive tower is approximately quantized in units of $\pi/z_r$.

In the quasi-static regimen of the dS scenario (\ref{g}); i.e., $\beta t\ll1$, the set of regularized states is used to determinate the gravitational potential between two massive particles separated by a distance $r$  on the physical brane.  Thus, we have
\begin{eqnarray}
V(r)&=&\frac{|\psi_0(0)|^2}{4\pi M_5^3}\frac{m_1m_2}{r}\left(1+\frac{4}{3\pi|\psi_0(0)|^2}\right.\nonumber\\
&&\times\left.\sum_{i=1}^{2}\int_{m_0}^\infty |\psi_m^i(0)|^2e^{-mr}z_r dm\right),\label{VN}
\end{eqnarray} 
where $M_5$ is the Planck mass in five-dimensions and $m_0=3\beta/2$.

\section{Resonance and Gravitational potential}\label{GraRes}

In analogy with the standard cosmology, $\beta$ in (\ref{g}) is the Hubble constant, around $10^{-9}$yr${}^{-1}$; and as a result, those configurations associated to a dS${}_5$ spacetime are similar to those scenarios embedded in a  five-dimensional flat bulk (remember that $\Lambda_\pm\leq6\beta^2$); where, unlike the AdS${}_5$ scenarios, the gravitational interaction on the brane has a five-dimensional behavior \cite{Kehagias:2002qk,Ghoroku:2003bs}. Thus, in order to consider the effects generated for the asymmetry on the Newtonian potential, we will focus in  the asymmetric cases with AdS${}_5$ vacua. 

\begin{figure*}[t]
\begin{minipage}[b]{0.33\linewidth}
\includegraphics[width=8cm,angle=0]{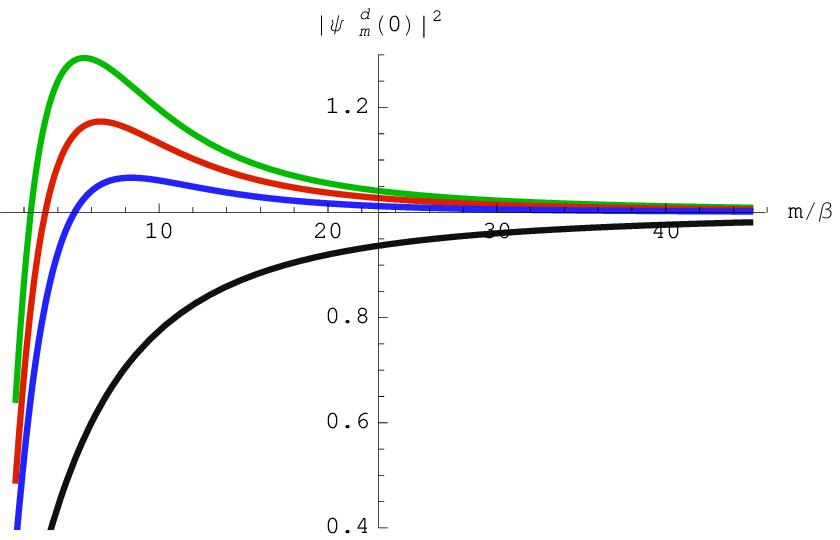}
\end{minipage}\hfill
\begin{minipage}[b]{0.45\linewidth}
\includegraphics[width=8cm,angle=0]{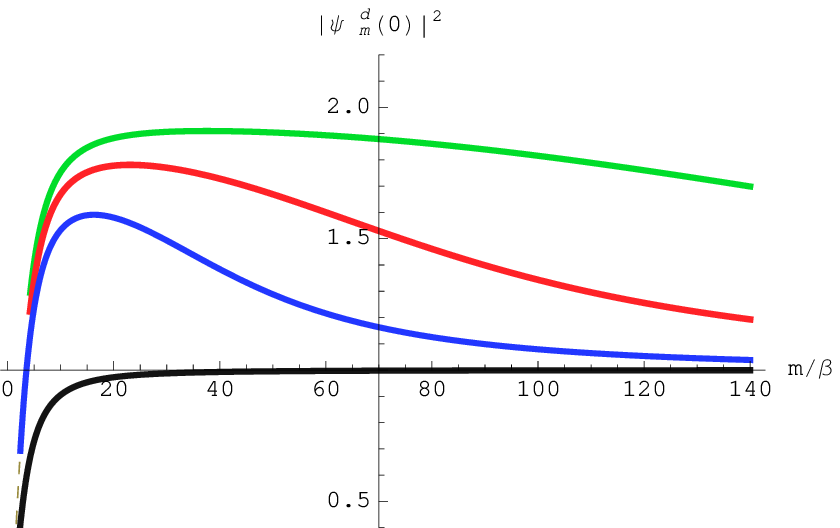}
\end{minipage}\hfill
\caption{ Resonant modes for different asymmetric AdS${}_5$ scenarios, $k_+/k_-=1\text{(black), 0.1\text{(blue)}, 0.05\text{(red)}, 0.01\text{(green)}}$. $k_+<m_0$ (left) and $k_\pm\gg m_0$ (right). }\label{reso}
\end{figure*}

About these scenarios, we must highlight the existence of resonant massive modes at $z=0$; i.e., massive states with a probability greater than others to be on the brane. In Fig.\ref{reso} the behavior of $\left|\psi^d_m(0)\right|^2$ for different  values of $k_+/k_-\leq 1$ is shown, where $k_\pm\equiv\sqrt{|\Lambda_\pm|/6}$; such that,  $k_+<m_0$  at left side of the figure,  while $k_\pm\gg m_0$ at right side of the figure. In accordance with Fig.\ref{reso}, in the first case, the resonant mode mass decreases when the asymmetry increases; unlike to the second  case, where the resonant mode mass increases for  large $k_+$ and $k_-$. In any case, it turns out that the resonant mass, for   $k_+/k_-\lesssim1/4$, is approximately given by  
\begin{equation}\label{mrAdsAds}
 m_{res}\sim \left[\left(k_+^2+ \frac{9}{4}m_0^2\right)\left(k_-^2+ \frac{9}{4}m_0^2\right)\right]^{1/4},
 \end{equation}
 such that it is bounded as follows 
 \begin{equation}
\frac{15}{4} k_+^2 +m_0^2< m^2_{res}< \frac{15}{4}k_-^2+m_0^2. \label{cotareso}
 \end{equation} 

 Notice that the upper bound and lower bound in (\ref{cotareso}), are given by $V_{QM}(0^-)$ and $V_{QM}(0^+)$, respectively. 
 
 Due to the exponential term in (\ref{VN}), we expected that the effective deviations to the Newton law are generated by the light  modes;  in such a way that,  the resonance contribution is negligible for those scenarios where the resonant mode belongs to the heavy states set. However, in those scenarios where the resonant mode is lightweight, its contribution to the gravitational potential (\ref{VN}) must be analyzed. In fact, as a consequence of the mass gap in (\ref{VQM}), in the scenarios where $k_+<m_0$, the increase of the asymmetry does not discourage the localization of the  zero mode and a Newtonian term in (\ref{VN}) is expected even when $k_+\rightarrow 0$.

 Next, we will determine the gravitational potential on three asymmetric AdS${}_5$ scenarios, where  the integral in (\ref{VN}) is always saturated at 
 \begin{equation}
 m_0\ll m\ll k_\pm. \label{cotam}
 \end{equation}

\subsection{ Weakly asymmetric AdS${}_5$ scenario}
 
Consider the almost symmetric scenario $k_+\sim k_-$, where the resonant mode is absent. So, to first order in $k_+/k_--1$, the integration constants are given by
\begin{eqnarray}
A_+&\simeq& -1+\frac{7}{3}\frac{m_0}{m}+\frac{7}{3}\frac{m}{k_+},\\
A_-&\simeq& -1+\frac{7}{3}\frac{m_0}{m}+\frac{7}{3}\frac{m}{k_+}-\frac{3}{2}\left(\frac{k_+}{k_-}-1\right),\\
B_-&=&1+\frac{3}{2}\left(\frac{k_+}{k_-}-1\right),
\end{eqnarray}
and the spectrum of the gravitational fluctuations is reduced to
 \begin{equation}
|\psi_0(0)|^2\simeq 2 \left(k_-^{-1}+k_+^{-1} \right)^{-1} \label{cero1}
\end{equation}
and 
\begin{equation}
|\psi^d_m(0)|^2\simeq \frac{4}{\pi}\left(\frac{k_+}{m}\right)^3\left[2-3\left(\frac{k_+}{k_-}-1\right)\right].\label{mass1}
\end{equation}

Substituting (\ref{cero1}) and (\ref{mass1})  in  (\ref{VN}), we have  
\begin{eqnarray}
V(r)&\simeq&\frac{k_+}{2\pi M_5^3}\left(\frac{k_+}{k_-}+1\right)^{-1} \frac{m_1m_2}{r}\left[1+ \frac{8}{3\pi^2}\left(\frac{k_+}{m_0}\right)^2\right.\nonumber\\
&&\times\left.\left(1-2m_0r\right)\left[2-3\left(\frac{k_+}{k_-}-1\right)\right]\right]\label{V(r)}
\end{eqnarray}
and Newtonian gravity is recovered on the brane when 
 \begin{equation}
\text{max}\{k_-^{-1}\ ,k_+^{-1}\}\ll r\ll  m_0^{-1}\ .
 \end{equation}

On the other hand, for $k_+\rightarrow k_- $, consistently  the Newtonian potential in the  $Z_2$ case is obtained \cite{Kehagias:2002qk,Ghoroku:2003bs}. 


\subsection{Strongly asymmetric AdS${}_5$ scenario}

Now, for large $k_+$ and $k_-$, but  $k_+\ll k_- $, the integration constants are determined as follows 
\begin{eqnarray}
A_+&\simeq&-1+\frac{1}{2}\frac{m_0}{m}+2 \frac{m}{k_+},\\
A_-&\simeq&\left(\frac{k_+}{k_-}\right)^{3/2}\left(-1+\frac{17}{12}\frac{m_0}{m}+\frac{m}{k_+}\right),\\
B_-&\simeq&\left(\frac{k_+}{k_-}\right)^{3/2}\left(1+\frac{11}{12}\frac{m_0}{m}-\frac{m}{k_+}\right),
\end{eqnarray}
and  the corresponding density of states are  given by  
\begin{equation}
|\psi_0(0)|^2\simeq \frac{k_+}{2} , \label{cero2}
\end{equation}
and 
\begin{equation}
z_r  |\psi^d_m(0)|^2\simeq\frac{16}{\pi} \left(\frac{k_+}{m}\right)^{3}.\label{mass2}
\end{equation}

After evaluating (\ref{VN}), we obtain 
\begin{eqnarray}
V(r)&\simeq&\frac{k_+}{8\pi M_5^3}\frac{m_1m_2}{r}\nonumber\\ 
&&\times\left[1+ \frac{64}{3\pi^2}\left(\frac{k_+}{m_0}\right)^2 \left(1-2 m_0 r\right) \right];\label{VAdSAdS}
\end{eqnarray}
and four-dimensional gravity is once again recovered, but in this case, in the region 
 \begin{equation}
 k_+^{-1}\ll r\ll m_0^{-1}\ .\label{cota2}
 \end{equation}
 
Now, comparing (\ref{cotareso}) and (\ref{cota2}), we find that the resonant mode is a heavy state whose contribution, along with the contribution of other heavy modes,  is suppressed for the exponential factor in (\ref{VN}).



\subsection{AdS${}_5$-Minkowski scenario}
In accordance with (\ref{cotareso}), to include the resonant mass within (\ref{cotam}) $k_+\rightarrow 0$ is required. In this case $\Lambda_+=0$, and from (\ref{N0}) we have 
\begin{eqnarray}
N_0^{-2}&=&\frac{1}{3\beta}-\frac{1}{2}\sqrt{\frac{6}{\Lambda_-}\left(\frac{6\beta^2}{\Lambda_-}-1\right)}
+\beta^2\sqrt{\left(\frac{6}{\Lambda_-}\right)^3}\nonumber\\ &&\times\tan^{-1}\left(\sqrt{\frac{6\beta^2}{\Lambda_-}}-\sqrt{\frac{6\beta^2}{\Lambda_-}-1}\right),
\end{eqnarray}
which is a positive and finite quantity  for  arbitrary $\Lambda_-$; and a bound zero mode is obtained  even when one of the portions of the five-dimension spacetime is flat, in correspondence with the mass gap in (\ref{VQMf2}). Additionally, notice that  in the static case
\begin{equation}
N_0^2(\beta\rightarrow 0)=2 i\left(\sqrt{\frac{6}{\Lambda_-}}+\sqrt{\frac{6}{\Lambda_+}}\right)^{-1},
\end{equation}
and only when the bulk is AdS${}_5$ the gravity is localized on the brane, as expected.

Let us analyze the effect of the light resonant mode on the gravitational potential (\ref{VN}).
 
For $k_+\rightarrow 0$, the integration constants are determined by
\begin{eqnarray}
A_+&\simeq&\frac{144}{33\pi}\left(\frac{k_-}{m}\right)^4\left(\frac{26}{11}\frac{m}{k_-}-\frac{m_0}{m}\right)\frac{m}{m_0} ,\\
A_-&\simeq&\frac{8}{33}\sqrt{\frac{1}{\pi}\left(\frac{k_-}{m}\right)^5}\left[\frac{1}{3}\left(6-7\frac{m_0}{m}\right)\frac{m_0}{m}\right.\nonumber\\
&&-\left.\frac{1}{11}\left(52+9\frac{m_0}{m}\right)\frac{m}{k_-}\right]\frac{m}{m_0}, \\
B_-&\simeq&\frac{8}{33}\sqrt{\frac{1}{\pi}\left(\frac{k_-}{m}\right)^5}\left[-\frac{1}{3}\left(6+7\frac{m_0}{m}\right)\frac{m_0}{m}\right.\nonumber\\
 &&+ \left.\frac{1}{11}\left(52-9\frac{m_0}{m}\right)\frac{m}{k_-}\right]\frac{m}{m_0},
\end{eqnarray}
 and it follows that 
 \begin{equation}
 |\psi_0(0)|^2\simeq 2 m_0,\quad z_r  |\psi^d_m(0)|^2\simeq 2.
\end{equation}
 
Thus, we find that the gravitational potential is given by
\begin{eqnarray}
V(r)\simeq\frac{m_0}{2\pi M_5^3}\frac{m_1m_2}{r} \left[1+ \frac{4}{3\pi} \frac{\left(1-m_0r\right)}{m_0r}\right];
\end{eqnarray}
such that, the contribution of the resonant mode is of the $1/r$ form  and five-dimensional  potential, $1/r^2$, appears as the dominant potential in the region 
\begin{equation}
k_-^{-1}\ll r \ll m_0^{-1}.
\end{equation}

Therefore,  when one of the vacua is null,  Newtonian gravity is not feasible  on the brane.

\section{Conclusions}\label{diss}

We analyzed the gravitational interaction on a  five-dimensional asymmetric dS spacetime. The configuration was generated by connecting rigidly at $z=0$ two vacuum solutions to the Einstein equations with planar-parallel  symmetry and different curvatures. As a result, on the four-dimensional  transition region a dS brane is obtained. Consistently, the spacetime has a null curvature limit equal to dS solution studied in \cite{Guerrero:2002ki}; and a static limit, with symmetric AdS${}_5$ vacua, similar to the RS-2 scenario \cite{Randall:1999vf}.  We focus on the scenarios with negative curvature, because, unlike the cases with curvature positive or zero,  the gravitational interaction on the brane is of the four-dimensional form \cite{Kehagias:2002qk,Ghoroku:2003bs}. 

It turns out that for a given asymmetric scenario, it is possible to find  a resonant mode inside of the spectrum of gravitational fluctuations. We noted that, in those scenarios where $k_+<m_0$, its mass decreases with the increases of asymmetry; while, in those scenarios characterized by $k_\pm\gg m_0$, its mass increases with the increases of asymmetry. In any case, the resonant mass remains bounded between the cosmological constants values at each side of the brane. 

In the AdS${}_5$ scenarios where $k_\pm\gg m_0$, the resonant mode belongs to the heavy states set and its contribution to the gravitational potential on the brane is exponentially suppressed, in such a way that Newtonian gravity is recovered on any four-dimensional sector of the scenarios. On the other hand, in order to include the resonant mode inside the light modes set, we explored the scenario AdS${}_5$-flat. In the dynamics case, although one of the vacua is null, the zero mode  of gravitational fluctuations is normalizable and an interaction Newtonian term is recovered in the gravitational potential. However, due to the resonant mode correction, five-dimensional gravity is obtained on the brane.

\section*{Acknowledgments}

RG and ROR wish to thank N. Pantoja for fruitful discussions and N. Romero for his collaboration to complete this paper. This work was supported by CDCHT-UCLA under project 018-CT-2009.

\newpage

\end{document}